# Universality in Intensity Modulated Photocurrent in Bulk-Heterojunction Polymer Solar Cells

Monojit Bag and K. S. Narayan*

Jawaharlal Nehru Centre for Advanced Scientific Research, Jakkur P. O.

Bangalore, 560064, India

**Abstract:** We observe a universal feature in the frequency ($\omega$) dependence of intensity modulated photocurrent $I_{ph}(\omega)$ based on studies of a variety of efficient bulk-heterojunction polymer solar cells (BHJ-PSCs). This feature of $I_{ph}(\omega)$ appears in the form of a local maximum in the 5 kHz < $\omega_{max}$ < 10 kHz range and is observed to be largely independent of the external parameters such as modulated light intensity ($L_{ac}$), wavelength ($\lambda$), temperature ($T$), and external field ($EF$) over a wide range. Simplistic kinetic models involving carrier generation, recombination and extraction processes are used to interpret the overall essential features of $I_{ph}(\omega)$ and correlate it to the device parameters.

______________________________

*Electronic mail: narayan@jncasr.ac.in

# I. INTRODUCTION

Bulk-heterojunction polymer solar cells (BHJ-PSCs) are promising alternatives for economic solution of renewable energy sources with relatively low manufacturing complexity. Efficiency ($\eta$) in the range of 5% has been achieved[1-3] for bulk-heterojunction polymer/fullerene solar cell. The general approach to improve these devices has been to explore novel compatible donor (D)-acceptor (A) systems which covers a wide absorption region[4,5] over the solar spectrum, along with minimizing losses in the different photophysical processes involving charge transfer and separation[6]. The network morphology[7,8] and the optimized gradients of the appropriate phases at the electrodes[9] in these BHJ-PSCs have also been observed to play an important role in determining the properties of these devices. There are growing options for the donor component of the BHJs, with a range of materials having desired energetic levels to promote the ultrafast photoinduced charge transfer processes[10,11]. A common set of features in these systems is the existence of ultrafast photoinduced charge transfer across the D-A interface, and open circuit voltage ($V_{OC}$) related to difference in the D-HOMO level and the A-LUMO level[12]. We observe an additional common feature in the BHJ-PSCs in the form of a local feature in modulated photocurrent ($I_{ph}$) and attribute it to a characteristic combination of correlated fundamental processes occurring over multiple time scales.

Intensity modulated photocurrent spectroscopy (IMPS) is a valuable complementary approach for transient measurements to study charge carrier-relaxation processes. This technique involves measuring the photocurrent response to a superimposition of small sinusoidal perturbation on larger dc illumination level. The use of relatively small modulation amplitudes has the advantage that the diffusion coefficient of electrons is primarily determined by the dc illumination intensity. This has been traditionally used for dye sensitized cells[13,14] and disordered a-Si cells[15,16] to highlight the relaxation mechanisms related to carrier life time, transit time and recombination dynamics. Phase shifts in modulated photocurrent (MPC) measurements in absence of a dc background have also been utilized to extract density of trap states[17,18]. We utilize a modified IMPS approach to understand the effect of various processes in BHJ-PSCs. In our approach, the primary source is ac-driven GaN based white LED light source [19 S1] using a function-generator, hence the sample is in dark state (during half the cycle). The additional background light intensity is introduced by a separate white CW LED light source driven by a constant voltage-source.

A large number of processes are initiated upon photoexcitation in BHJ-PSCs spanning over a large range of time scales. A steady state response for modulated input is a net outcome of these varied processes at that frequency ($\omega$). In this context, it is interesting to observe a unified $I_{ph}(\omega)$ response prevalent in large variety of PSCs which is relatively independent of the intrinsic material constituents, interfacial characteristics such as energy level differences and external factors such as temperature ($T$) and modulated light intensity ($L_{ac}$). We demonstrate that these results, which encompass a restricted frequency range, can be explained by a solution to a simplified kinetic model which takes into account only the primary processes of carrier generation, recombination, relaxation and extraction. $I_{ph}(\omega)$ is observed to be independent of $\omega$ for $\omega < 1$ kHz unless there are slow relaxation processes associated with factors such as space-charge build-up. We highlight the prominent feature in the mid-frequency regime (5 kHz $< \omega_{max} <$ 10 kHz) in a form of a local maximum in $I_{ph}(\omega)$ which is observed in all efficient BHJ-PSCs. To understand $I_{ph}(\omega)$ profile of these devices, we have used kinetic models which also enable us to correlate $\eta$, fill factor ($FF$) and other important device parameters.

## II. EXPERIMENT

Devices were fabricated using optimized protocol to obtain devices with appropriate morphology[20,21], modifying polymer:electrode interface, and size of the device/electrode. Poly(3,4-ethylenedioxythiophene): poly(styrenesulfonic acid) (PEDOT:PSS) (Baytron P) is spin coated (~ 80 nm) on a pre-cleaned ITO coated (~ 20 Ω/□) glass substrate and annealed at 120°C for 30 min. Donor-acceptor BHJ blend in chlorobenzene solvent (20 mg/ml) is spin coated (~ 100 nm) on PEDOT:PSS layer and cured at 80°C over which Al (cathode) is thermally deposited at $10^{-6}$ mbar. All the devices exhibited $\eta$ of at least > 1 %, which was measured using calibrated light source. The absence of contributions from possible frequency dependence of the light source output and the associated measurement-electronics in the set up was rigorously verified [see S2].

Figure 1 represents a typical $I_{ph}(\omega)$ response from BHJ PSC with $\eta$ > 1 %. The set of BHJ:PSCs were based on poly-[3-hexylthiophene] (P3HT) and poly[2,6-(4,4-bis-(2-ethylhexyl)-4H-cyclopenta[2,1-b;3,4-b']-dithiophene)-alt-4,7-(2,1,3-benzothiadiazole)] (PCPDTBT) donor polymer[22] blended with acceptor either [6,6]-phenyl $C_{61}$ butyric acid methyl ester (PCBM) macromolecule[23] or poly{[N,N'-bis(2-octyldodecyl)-naphthalene-1,4,5,8-bis(dicarboximide)-2,6-diyl]-alt-5,5'-(2,2'-bithiophene)} (P(NDI2OD-T2), Polyera ActivInk N2200) polymer[24]. The typical value of $I_{SC}$ (responsivity) and $V_{OC}$ were in the range of 20-30 mA/W and 0.55-0.6 V respectively in all the solar cells studied.

As the light source ($L_{ac}$ ~ 1 mW/cm$^2$) incident on the PSCs is modulated, $I_{ph}(\omega)$ magnitude marginally changes from the $\omega = 0$ value and is relatively ω-independent at low frequencies ($\omega$ < 1 kHz). For $\omega$ > 1 kHz, $I_{ph}(\omega)$ increases with $\omega$ and a distinct maximum of $I_{ph}(\omega_{max})$ is observed in the range of 5 kHz < $\omega_{max}$ < 10 kHz. Beyond 10 kHz, $I_{ph}(\omega)$ decreases and follows a power-law decay with an exponent γ ~ 2. The phase ($\phi$) of a typical $I_{ph}(\omega)$ response indicates a singularity ($d\phi(\omega)/d\omega|_{\omega=\omega_{max}} = -\infty$) at $\omega_{max}$. The cole-cole representation of this data set reveals this feature in the form of a distorted circle. The magnitude of $I_{ph}(\omega)$ scales with probe light intensity and the wavelength ($\lambda$), but the $\omega$ dependence with respect to these external parameters is preserved over a wide range for all the efficient devices tested in our laboratory. Possibilities of frequency dependence of the light sources have also been checked

## III. MODEL

The uniformity of the response for all the combinations of BHJ-PSCs studied points to a set of common dominant underlying processes in this class of systems. It is instructive to model the response as a product of two independent transfer functions consisting of a circuit component and a microscopic component. The circuit model consists of voltage dependent dynamic resistance (represented as a diode) with a current source ($I_L$) along with two parasitic resistances as shown in the figure 2(a). $R_{sh}$ represents the leakage loss whereas the voltage dependent $R_s$ represents the charge extraction resistance[25,26]. The circuit transfer function (CTF) takes the form of ~ $\frac{I_{ph}(\omega)}{I_L(\omega)} = \frac{R_{sh}}{[R_{sh}R_s C_p i\omega + (R_{sh}+R_s)]}$ where $I_{ph}(\omega)$ is a measure of short circuit current. The effective parallel capacitance $C_p$ is the combination of bias-dependent junction capacitance ($C_T$) and voltage- and frequency-dependent diffusion capacitance ($C_D$)[27]. $C_p$ can be expressed as $C/(ai\omega+1)$ where $C$ represents the geometrical factor and '$a$' is governed by intrinsic factors. The form of this CTF can then be further reduced to ~ $q(ai\omega + 1)/[(a + qR_sC)i\omega + 1]$ where $q = R_{sh}/(R_{sh} + R_s)$. The basic prevailing microscopic kinetic processes of carrier generation, recombination and charge carrier trapping along with associated rate constants [Fig. 2b] form the microscopic $I_L(\omega)$. These microscopic processes span over multiple time scales (pico - millisecond). The associated parameters are primary excitons generation rate ($g_L$), exciton diffusion lifetime ($\tau_1 \sim$ ns), charge-carrier ($n_{free}$) life time $\tau_2$, secondary exciton generation process, and trapping kinetics[28]. A simplified kinetic form of the processes which involve exciton dissociation rate constant ($K_{diss}$), free carrier recombination rate constant ($K_{recom}$) and effective charge carrier trapping rate constant ($\kappa_{eff}$) then can be expressed as

$$K_1 \frac{d^2 n_{free}(t)}{dt^2} + K_2 \frac{d n_{free}(t)}{dt} + n_{free}(t) = -Q g_L(t) \quad \ldots\ldots\ldots\ldots\ldots\ldots\ldots\ldots(1)$$

and the solution in Fourier space takes the form $N_{free}(\omega) = \frac{-Q}{[-K_1\omega^2 + K_2 i\omega + 1]} G_L(\omega)$, where $K_1 = 1/(\kappa_{eff} - K_{recom} K_{diss})$, $K_2 = 1/\{\tau_2(\kappa_{eff} - K_{recom} K_{diss})\}$ and $Q = K_{diss}/\{\tau_1(\kappa_{eff} - K_{recom} K_{diss})\}$. Generation current $I_L(\omega)$ is proportional to the number of free carriers $N_{free}(\omega)$ at a given rate of charge extraction. The over-all transfer function (TF) then takes the form $Pq \frac{(ai\omega + 1)}{[\{(a + qR_sC)i\omega + 1\}(-K_1\omega^2 + K_2 i\omega + 1)]}$ where $P$ is a device constant that depends on carrier mobility and exciton life time. A detail of the model equation is given in appendix A.

### IV. RESULTS AND DISCUSSION

This expression for TF is utilized to fit the experimental results assuming a set of realistic constants [Fig. 3]. The model fit relies essentially on the two parameters ($K_1$ and $K_2$) describing the Lorentzian in the denominator and is independent of other variables [Table 1]. It is noticed that the $\omega_{max}$ ($\approx K_1^{-1/2}$) feature is less sensitive to the circuit parameters $R_s$, $R_{sh}$ and $C_p$. It should be noted that the proposed model [Eq. 1] can be further validated by solving it in the time domain and comparing it with the experimentally obtained transient $I_{ph}(t)$ for the same set of parameters ($K_1$ and $K_2$)[29]. The frequency dependent magnitude and phase of the transfer function also agree reasonably well with the experimentally obtained $I_{ph}(\omega)$ magnitude and the corrected phase ($\phi$).

#### A. Temperature dependent study

Figure 4 depicts the observed temperature dependence of $I_{ph}(\omega)$. $I_{ph}(\omega)$ decreases in an activated manner as expected, however the value of the $\omega_{max}$ in the response profile remains the same over the entire $T$ range. This characteristic behavior of $\omega_{max}$ with respect to $T$ and modulated light intensity indicates the absence of a single microscopic process as a decisive factor in controlling $\omega_{max}$. If the response originates exclusively from the long-lived relaxation mechanisms related to trap level kinetics then a shift in $\omega_{max}$ would be expected as a function of temperature and intensity. This interpretation is also consistent with the constancy of the results for different set of BHJ-interfaces where one expects a large variation in the trap energetics and distribution.

The representation of the complex frequency response largely as Lorentzian transfer function identifies the source for the maximum in $I_{ph}(\omega)$. The analogy with a damped driven oscillator system where the transfer function has a similar form is useful. Photocarrier generation can be taken as a driving term whereas the trapping and recombination terms are the loss mechanisms of the system. Another possible scenario is that the invariant $\omega_{max}$ feature can originate from the dimensions, organization and coupling of the nanoscale phase separated regions[19]. Geometrical factors as a source for a characteristic $\omega$ then can also discount the dependence on the external factors.

The circuit component of the transfer function can be used in examining the $T$ dependence closely in the low $\omega$ region ($\omega$ <1 kHz). In this range $I_{ph}(\omega)$ is dependent on $T$ with $dI_{ph}(\omega)/d\omega < 0$ below room temperature. A trap limited transport model can be used to describe the low $\omega$ region ($\omega$ <1 kHz), with trap states exponentially distributed below the band edge. Temperature dependent $R_s$ can take a form $R_s = R_0 \exp(E_a/k_B T)$ where $E_a$ is the activation energy for the free carriers at the polymer electrode interface. $R_s$ increases from 5 k$\Omega$ to 22 k$\Omega$ as temperature is lowered from 250 K to 120 K [Table 2]. Activation energy $E_a$ is estimated ~ 30 meV from the $R_s$-$T^{-1}$ plot [Fig. 4 inset]. The $I_{ph}(\omega)$ trend is also observed to be correlated to the device characteristics. For instance, it is observed that $dI_{ph}(\omega)/d\omega > 0$ for low $FF$ devices [Fig. 5]. This trend was verified for a large number of low $FF$ devices which can be realized by skewed composition (D:A) ratio, incomplete annealing process, and cathode-BHJ effects[21].

### B. Effect of background light

Another interesting parameter in these studies is the dc-background light intensity ($L_{dc}$). The presence of dc background light results in a red shift of $\omega_{max}$, with $\omega_{max}$ decreasing exponentially with respect to $L_{dc}$. As $L_{dc}$ is increased to the probe intensity level of ~ 1 mW/cm$^2$, $\omega_{max}$ decreases from 5.7 kHz to 4.5 kHz and is accompanied by a broadening of the photocurrent peak (Fig. 6). The dc-excitation alters the trap-site occupancy and increases the carrier density in the trap states ($n_t$) along with an increase in the recombination rate leading to a decrease in $I_{ph}(\omega)$. An effective increase in $K_1$ can be physically justified as the $K_{recom}$ term increases as well as $\kappa_{eff}$ decreases [Table 3], leading to a decrease in $\omega_{max}$ as per the expression for $I_{ph}(\omega)$ derived from the microscopic model.

### V. MODULATED PHOTOCURRENT FROM OTHER DEVICES

Intensity modulated photocurrent measurements were additionally carried out for comparison with the dye-sensitized solar cells (DSSC) as well as bi-layer polymer solar cells where the nature of charge carrier transport and associated dynamics are different from that in BHJ-PSCs. In case of DSSCs, charge transfer occurs is initiated at the TiO$_2$/dye interface and electron transport is largely governed by the TiO$_2$ nano-structure layer characteristics, while the ionic conduction through the electrolyte depends on factors including ion concentration and ionic-mobility. IMPS measurements on DSSCs have been extensively studied and our measurements on such cells are quite similar to the reported results [30].

In case of bi-layer organic solar cells such as ITO/P3HT/PCBM/Al, charge separation occurs primarily at the interface and the electron (hole) is transported through n-type (p-type) polymer chain respectively (unipolar transport) [31]. The kinetic equation representing the transport is expected to be different. Figure 7 represents the modulated photocurrent features of such three different types of polymer solar cells.

## VI. CONCLUSION

$I_{ph}(\omega)$ response from BHJ-PSCs is a compact representation of the various photophysical and electrical processes prevalent in these systems. The equivalence of the routes which culminate in the measured $I_{ph}(\omega)$ across different systems is clearly evident. This unifying $I_{ph}(\omega)$ response appears in well defined frequency regime. It was also possible to correlate $I_{ph}(\omega)$ in the low frequency regime to device limiting performance parameters.

## ACKNOWLEDGEMENTS

We thank Dr. N. Vidhyadhiraja for valuable discussions and Department of Atomic Energy, Government of India for partial funding.

### Appendix A: Calculation of generation current $I_L$:

Photon absorption occurs at femto-second scale. So it is almost instantaneous and internal quantum efficiency ~ 100%. However decay of photo generated excitons $g_L(t)$ depends on two processes. Excitons get dissociate at the D–A interfaces, and the rate at which it dissociates is inversely proportional to the average life time of the excitons and proportional to the no. of excitons at any instant. At the same time it can be generated from the free carriers $n_{free}(t)$ at a rate constant $K_{recom}$. Assuming average life time of the excitons $\tau_1$ we can write decay equation in the form.

$$\frac{dg_L(t)}{dt} = -\frac{g_L(t)}{\tau_1} + K_{recom} n_{free}(t) \quad \ldots\ldots\ldots\ldots\ldots\ldots\ldots\ldots\ldots (1)$$

Free charge carrier ($n_{free}$) will decay through several pathways. One of them is the charge carrier trapping at the deep trap states (much greater than thermal energy). There will be a significant de-trapping. However over all loss of carrier can be given by the equation $Q_{trap}\big|_{t=t_0} = \int_{t_0-\Delta t}^{t_0+\Delta t} \kappa_{eff} n_{free}(t) dt$ where $\kappa_{eff}$ is the effective trapping rate. Similarly we can write the rate equation of free carriers as

$$\frac{dn_{free}(t)}{dt} = -\frac{n_{free}}{\tau_2} + K_{diss} g_L(t) - Q_{trap} \quad \ldots\ldots\ldots\ldots\ldots\ldots\ldots (2.a)$$

Or, $$\frac{dn_{free}(t)}{dt} = -\frac{n_{free}}{\tau_2} + K_{diss} g_L(t) - \int \kappa_{eff} n_{free}(t) dt \quad \ldots\ldots\ldots\ldots (2.b)$$

Where $K_{diss}$ is the excitons dissociation rate constant and $\tau_2$ is the life time of the free carriers.

Combining two equations the relation between $g_L(t)$ and $n_{free}(t)$ could be written as

$$\frac{d^2 n_{free}(t)}{dt^2} + \frac{1}{\tau_2}\frac{dn_{free}(t)}{dt}(\kappa_{eff} - K_{recom} K_{diss}) n_{free}(t) = -\frac{K_{diss}}{\tau_1} g_L(t) \ldots (3.a)$$

Or, $K_1 \dfrac{d^2 n_{free}(t)}{dt^2} + K_2 \dfrac{dn_{free}(t)}{dt} + n_{free}(t) = -Q g_L(t)$ ……................... (3.b)

Where $K_1 = \dfrac{1}{(\kappa_{eff} - K_{recom} K_{diss})}$, $K_2 = \dfrac{1}{\tau_2 (\kappa_{eff} - K_{recom} K_{diss})}$ and

$Q = \dfrac{K_{diss}}{\tau_1 (\kappa_{eff} - K_{recom} K_{diss})}$

Equation 3 could be rewritten in frequency domain ($\omega$) after taking Fourier Transformation

$N_{free}(\omega) = \left[ \dfrac{-Q}{-K_1 \omega^2 + K_2 i\omega + 1} \right] G_L(\omega)$ ……………….……………….. (4)

Photocurrent [$I_L(\omega) \propto N_{free}(\omega)$] would be given by the equation

$I_L(\omega) = -P \left[ \dfrac{1}{-K_1 \omega^2 + K_2 i\omega + 1} \right] G_L(\omega)$ ……………………………….. (5)

*P* is a constant and it depends on mobility of the carriers, life time of the excitons and dissociation rate constant. So the transfer function will be

$TF = -P \dfrac{1}{[-K_1 \omega^2 + K_2 i\omega + 1]}$ …………………………………………………. (6)

$I_{ph}(\omega) = TF_{macroscopic} I_L(\omega) = TF_{macroscopic} \times TF_{microscopic} G_L(\omega)$ ……................. (7)

So, $\dfrac{I_{ph}(\omega)}{G_L(\omega)} = TF_{total} = TF_{macroscopic} \times TF_{microscopic}$ ……………………………… (8)

**References:**

1. F. Padinger, R. S. Rittberger, and N. S. Sariciftci Adv. Funct. Mater. **13**, 85 (2003).

2. M. Reyes-Reyes, K. Kim, and D. L. Carroll, Appl. Phys. Lett. **87**, 083506 (2005).

3. G. Li, V. Shrotriya, J. Huang, Y. Yao, T. Moriatry, K. Emery, and Y. Yang, Nature Mater. **4**, 864 (2005).

4. F. Zhang, E. Perzon, X. Wang, W. Mammo, M. R. Andersson and Olle Inganäs, Adv. Func. Mater. **15**, 745 (2005).

5. N. Blouin, A. Michaud, and Mario Leclerc, Adv. Mater. **19**, 2295 (2007).

6. T. M. Clarke, F. C. Jamieson, and J. R. Durrant, J. Phys. Chem. C, **113**, 20934 (2009).

7. C. R. McNeill, A. Abrusci, I. Hwang, M. A. Ruderer, P. Müller-Buschbaum, and N. C. Greenham, Adv. Funct. Mater. **19**, 3103 (2009).

8.  W. Ma, C. Yang. X. Gong, K. Lee and A. J. Heeger, Adv. Funct. Mater. **15**, 1617 (2005).

9.  S. S. van Bavel, E. Sourty, G. de With and J. Loos, Nano Lett. **9**, 507 (2009).

10. N. S. Sariciftci, L. Smiloqitz, A. J. Heeger and F. Wudl, Science **258**, 1474 (1992).

11. G. Yu, J. Gao, J. C. Hummeler, F. Wudl and A. J. Heeger, Science **270**, 1789 (1995).

12. J. Cremer, P. Baüerle, M. M. Wienk, and R. A. J. Janssen, *Chem. Mater.* **18***, 5832 (2006).

13. F. F-Santiago, J. Bisquert, E. Palomeres, S. A. Haque, and J. R. Durrant, J. of Appl. Phys. **100**, 034510 (2006).

14. G. Schlichthörl, S. Y. Huang, J. Sprague, and A. J. Frank, J. Phys. Chem. B, **101**, 8141 (1997).

15. D. Vanmaekelbergh, J. van de Lagemaat and R. E. I. Schropp, Sol. Energy Mater. & Solar Cells **41-42**, 537 (1996).

16. I. Mora-Seró, Y. Luo, G. Garcia-Belmonte, J. Bisquert, D. Muñoz, C. Vozb, J. Puigdollers and R. Alcubilla, Sol. Energy Mater. & Solar Cells **92**, 505 (2008).

17. J. Orenstein and M. Kastner, Phys. Rev. Lett. **46**, 1421 (1981).

18. T. Agostinelli, M. Caironi, D. Natali, M. Sampietro, M. Arca, F. A. Devillanova and V. Ferrero, Synthetic Metals **157**, 984 (2007).

19. Supplementary information.

20. K. Kim, J. Liu, M. A. G. Namboothiry, and D. L. Carroll, Appl. Phys. Lett. **90**, 163511 (2007).

21. D. Gupta, M. Bag and K. S. Narayan, Appl. Phys. Lett. **92**, 093301 (2008).

22. J. Peet, J. Y. Kim, N. E. Coates, W. L. Ma, D. Moses, A. J. Heeger and G. C. Bazana, Nature Matter. **6**, 497 (2007).

23. S. Cook, H. Ohkita, Y. Kim, J. J. Benson-Smith, D. D. C. Bradley, J. R. Durrant, Chem. Phys. Lett. **445**, 276 (2007).

24. H. Yan, Z. Chen, Y. Zheng, C. Newman, J. R. Quinn, F. Dötz, M. Kastler and A. Facchetti, Nature **457**, 679 (2009).

25. S. Yoo, B. Domercq, and B. Kippelen, J. Appl. Phys. **97**, 103706 (2005).

26. B. Mazhari, Sol. Energy Mater. & Solar Cells, **90**, 1021 (2006).

27. H.S. Rauschenbach, Solar cell array design handbook, Van Nostrand Reinhold, (1980).

28. P. Kounavis, Phys. Rev. B, **64**, 045204 (2001).

29. The complete transfer function is solved in time domain for a set of parameters used for modulated photocurrent measurement and the solution (in the micro - millisecond scale) of $I_{ph}(t)$ yields good fit to the measured transient photocurrent.

30. L. Dloczik, O. Ileperuma, I. Lauermann, L. M. Peter, E. A. Ponomarev, G. Redmond, N. J. Shaw and I. Uhlendorf, J. Phys. Chem. B, **101**, 10281 (1997).

31. M. Koehler, L. S. Roman, O. Inganäs and M. G. E. da Luz, J. Appl. Phys. **96**, 40 (2004).

**Figure Caption:**

**Figure 1.** Intensity modulated photocurrent spectrum from 10 Hz to 100 kHz for PCPDTBT:PCBM (■), PCPDTBT:N2200 (●), P3HT:N2200 (▲) and P3HT:PCBM (▼) with a white LED illumination (GaN) driven by a frequency generator.

**Figure 2. (a).** Equivalent circuit diagram of a solar cell. $R_s$ and $R_{sh}$ are two parasitic resistances arising from charge extraction process and leakage current **(b)** Schematic diagram of microscopic processes involved in an organic solar cell. $K_{diss}$ and $K_{recom}$ are the exciton dissociation rate constant and free carrier recombination rate constant. $\kappa_{eff}$ represents effective rate of change of trapping rate constant of the charge carriers to thermally inaccessible states (>>25 meV). $G_L$ and $N_{free}$ represent total number of excitons and free carriers at any instant of time.

**Figure 3:** Modulated photocurrent amplitude (□) and phase (○) plot of a P3HT:PCBM devices. Simulation (black line for photocurrent and red line for phase) data matches with the experimental values for the following parameters and constants: $K_1 = 2.8 \times 10^{-8}$, $K_2 = 9 \times 10^{-5}$, $R_{sh} = 50$ kΩ, $R_s = 800$ Ω, $C = 38$ nF, $a = 1.6 \times 10^{-3}$.

**Figure 4.** Temperature dependent modulated photocurrent of a P3HT:PCBM blend solar cell. Symbols represent experimental values whereas line (with color) represents simulated photocurrent for a set of constant. Parameters varies with temperature are $R_s$ and $\tau_2$. Inset shows temperature dependent $R_s$ ($R_s = R_0 \exp[E_a/k_B T]$). Estimated activation energy ($E_a$) is 30 meV.

**Figure 5**. Modulated photocurrent of N2200:P3HT and N2200:PCPDTBT. Inset: light I-V curve for N2200:P3HT (□) and N2200:PCPDTBT (○) showing *FF* 0.51 and 0.16 respectively.

**Figure 6:** Modulated photocurrent with (○) and without (□) dc background light keeping probe intensity constant. Red (lower $\kappa_{eff}$ and higher $K_{recom}$) and black (higher $\kappa_{eff}$ and lower $K_{recom}$) line for simulation keeping other parameters constant.

**Figure 7:** Photocurrent comparison of three devices: Intensity modulated photocurrent spectrum of a BHJ-PSC (■), Modulated photocurrent of a dye-sensitized solar cell (▲) made of TiO$_2$ nano-porous structure coated with N3 dye and I$_2$ based quasi solid state electrolyte and modulate photocurrent spectrum of a bi-layer PSC (●) made of P3HT layer spin casted on ITO glass (20 mg/ ml concentration in Chlorobenzene solvent) and PCBM layer on top (20 mg/ml concentration in dichloromethane).

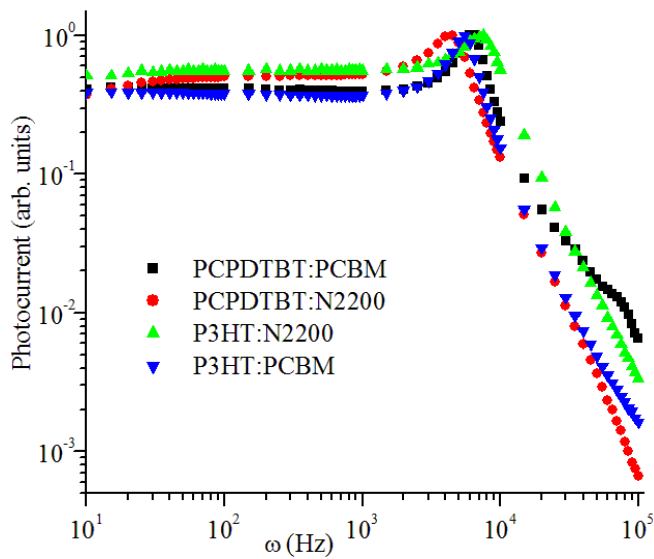

Figure 1

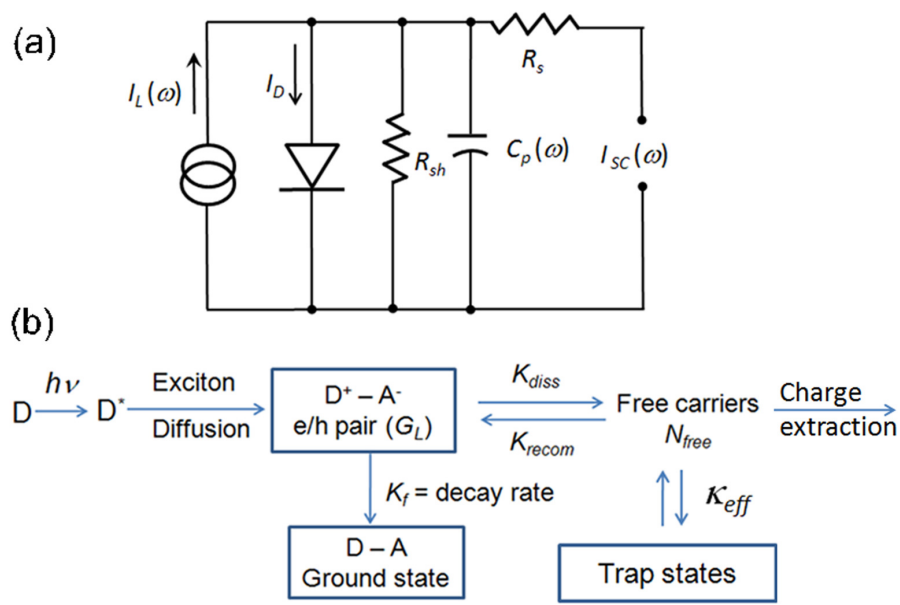

Figure 2

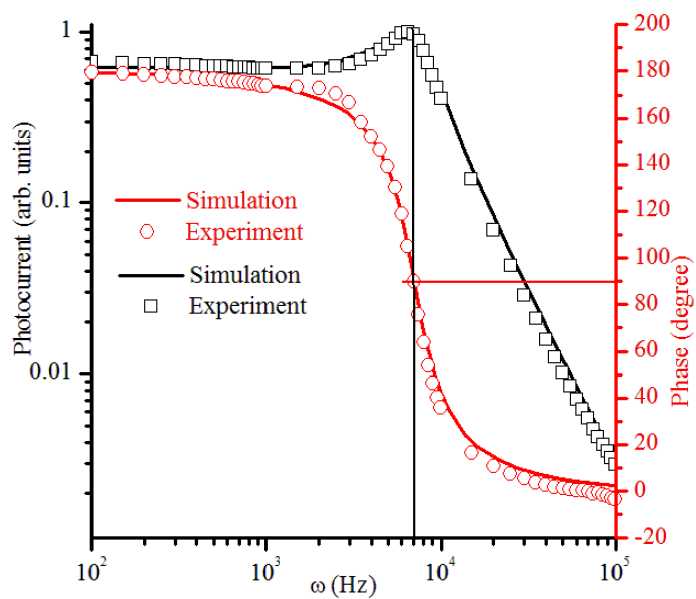

Figure 3

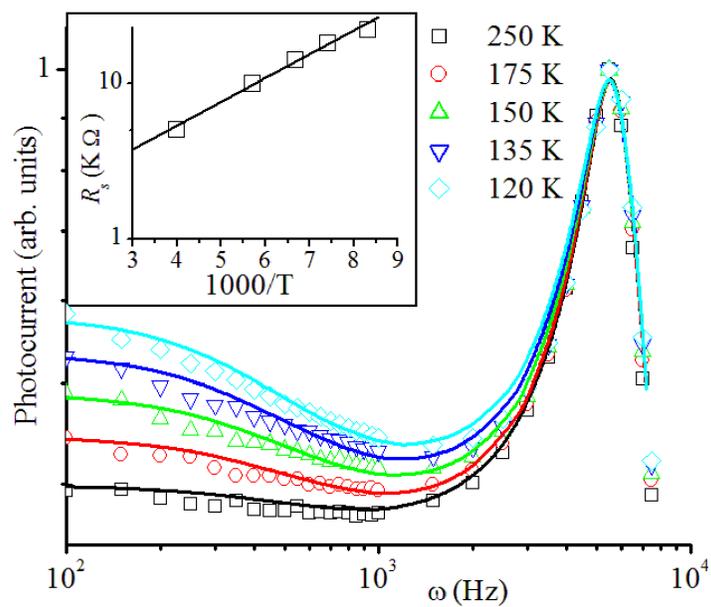

Figure 4

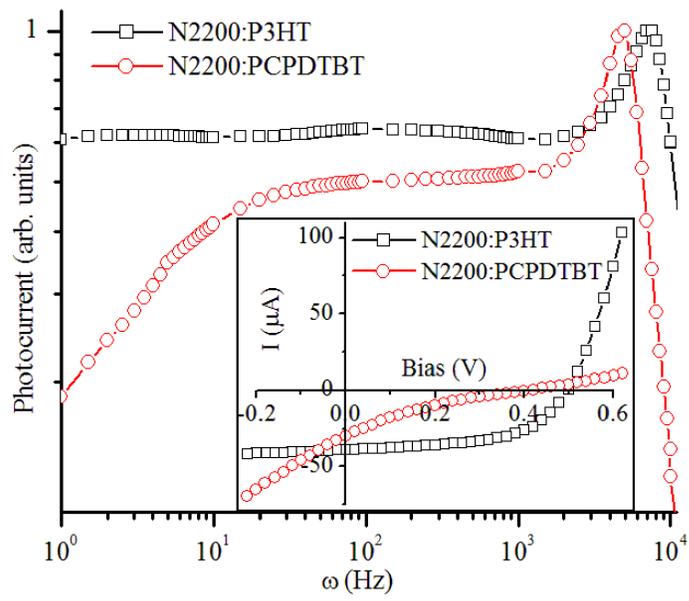

Figure 5

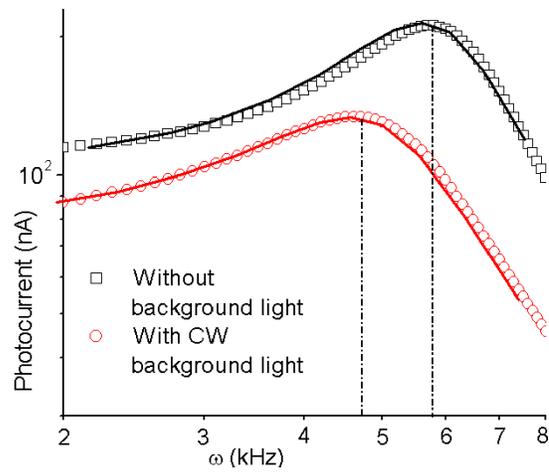

Figure 6

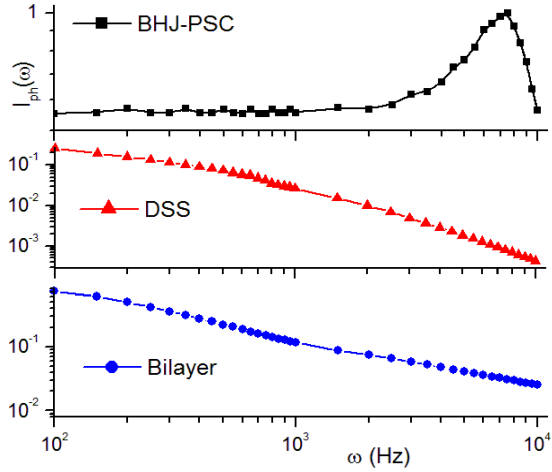

Figure 7

**Table 1:** Parameters used for the simulation of a P3HT:PCBM blend (1:1 by weight ratio) solar cell modulated photocurrent.

| Parameters | $R_s$ | $R_{sh}$ | $C_p$ | $a$ | Fit parameter ($K_1$) | Fit parameter ($K_2$) |
|---|---|---|---|---|---|---|
| Value | 800 Ω | 50 kΩ | 38 nF | 0.0016 | $2 \times 10^{-8}$ | $9 \times 10^{-5}$ |

**Table 2:** Parameter used for temperature dependent photocurrent fit of P3HT:PCBM device, $R_{sh}$ = 200 kΩ, $C_p$ = 38 nF, $a$ = 0.0016 and $K_1$ = $3 \times 10^{-8}$.

| Temperature | 250 K | 175 K | 150 K | 135 K | 120 K |
|---|---|---|---|---|---|
| $R_s$ | 5 kΩ | 10 kΩ | 14 kΩ | 18 kΩ | 22 kΩ |
| Fit parameter ($K_2$) | $6.3 \times 10^{-5}$ | $6.4 \times 10^{-5}$ | $6.6 \times 10^{-5}$ | $6.8 \times 10^{-5}$ | $7.0 \times 10^{-5}$ |
| $\tau_2$ (estimated) | 0.48 ms | 0.47 ms | 0.45 ms | 0.44 ms | 0.43 ms |

**Table 3:** Parameter used for different background light dependence photocurrent fit of P3HT:PCBM device, $R_s$ = 800 Ω, $R_{sh}$ = 50 kΩ, $C_p$ = 38 nF, $a$ = 0.0016.

| Parameters | No background light | With background light |
|---|---|---|
| $K_1$ | $2.8 \times 10^{-8}$ | $4.0 \times 10^{-8}$ |
| $K_2$ | $8.2 \times 10^{-5}$ | $12 \times 10^{-5}$ |
| $\tau_2$ | 0.34 ms | 0.33 ms |

13